\documentclass[twocolumn,showpacs,preprintnumbers,amsmath,amssymb,aps,prl,superscriptaddress]{revtex4}

\usepackage{graphicx}
\usepackage{dcolumn}
\usepackage{bm}

\begin{document}

\preprint{}

\title{Cross Sections and Transverse Single-Spin Asymmetries in Forward 
Neutral Pion Production from Proton Collisions at $\mathbf{\sqrt{s}=200}$ 
GeV}

%
\affiliation{Argonne National Laboratory, Argonne, Illinois 60439}
\affiliation{Brookhaven National Laboratory, Upton, New York 11973}
\affiliation{University of Birmingham, Birmingham, United Kingdom}
\affiliation{University of California, Berkeley, California 94720}
\affiliation{University of California, Davis, California 95616}
\affiliation{University of California, Los Angeles, California 90095}
\affiliation{Carnegie Mellon University, Pittsburgh, Pennsylvania 15213}
\affiliation{Creighton University, Omaha, Nebraska 68178}
\affiliation{Nuclear Physics Institute AS CR, \v{R}e\v{z}/Prague, Czech Republic}
\affiliation{Laboratory for High Energy (JINR), Dubna, Russia}
\affiliation{Particle Physics Laboratory (JINR), Dubna, Russia}
\affiliation{University of Frankfurt, Frankfurt, Germany}
\affiliation{Indiana University, Bloomington, Indiana 47408}
\affiliation{Institute  of Physics, Bhubaneswar 751005, India}
\affiliation{Institut de Recherches Subatomiques, Strasbourg, France}
\affiliation{University of Jammu, Jammu 180001, India}
\affiliation{Kent State University, Kent, Ohio 44242}
\affiliation{Lawrence Berkeley National Laboratory, Berkeley, California 94720}\affiliation{Max-Planck-Institut f\"ur Physik, Munich, Germany}
\affiliation{Michigan State University, East Lansing, Michigan 48824}
\affiliation{Moscow Engineering Physics Institute, Moscow Russia}
\affiliation{City College of New York, New York City, New York 10031}
\affiliation{NIKHEF, Amsterdam, The Netherlands}
\affiliation{Ohio State University, Columbus, Ohio 43210}
\affiliation{Panjab University, Chandigarh 160014, India}
\affiliation{Pennsylvania State University, University Park, Pennsylvania 16802}
\affiliation{Institute of High Energy Physics, Protvino, Russia}
\affiliation{Purdue University, West Lafayette, Indiana 47907}
\affiliation{University of Rajasthan, Jaipur 302004, India}
\affiliation{Rice University, Houston, Texas 77251}
\affiliation{Universidade de Sao Paulo, Sao Paulo, Brazil}
\affiliation{University of Science \& Technology of China, Anhui 230027, China}
\affiliation{Shanghai Institute of Nuclear Research, Shanghai 201800, P.R. China}
\affiliation{SUBATECH, Nantes, France}
\affiliation{Texas A\&M, College Station, Texas 77843}
\affiliation{University of Texas, Austin, Texas 78712}
\affiliation{Valparaiso University, Valparaiso, Indiana 46383}
\affiliation{Variable Energy Cyclotron Centre, Kolkata 700064, India}
\affiliation{Warsaw University of Technology, Warsaw, Poland}
\affiliation{University of Washington, Seattle, Washington 98195}
\affiliation{Wayne State University, Detroit, Michigan 48201}
\affiliation{Institute of Particle Physics, CCNU (HZNU), Wuhan, 430079 China}
\affiliation{Yale University, New Haven, Connecticut 06520}
\affiliation{University of Zagreb, Zagreb, HR-10002, Croatia}
\author{J.~Adams}\affiliation{University of Birmingham, Birmingham, United Kingdom}
\author{C.~Adler}\affiliation{University of Frankfurt, Frankfurt, Germany}
\author{M.M.~Aggarwal}\affiliation{Panjab University, Chandigarh 160014, India}
\author{Z.~Ahammed}\affiliation{Variable Energy Cyclotron Centre, Kolkata 700064, India}
\author{J.~Amonett}\affiliation{Kent State University, Kent, Ohio 44242}
\author{B.D.~Anderson}\affiliation{Kent State University, Kent, Ohio 44242}
\author{M.~Anderson}\affiliation{University of California, Davis, California 95616}
\author{D.~Arkhipkin}\affiliation{Particle Physics Laboratory (JINR), Dubna, Russia}
\author{G.S.~Averichev}\affiliation{Laboratory for High Energy (JINR), Dubna, Russia}
\author{S.K.~Badyal}\affiliation{University of Jammu, Jammu 180001, India}
% ...

\collaboration{STAR Collaboration}\homepage{www.star.bnl.gov}\noaffiliation

\date{\today}

\begin{abstract}
Measurements of the production of forward high energy $\pi^0$ mesons 
from transversely polarized proton collisions at $\sqrt{s}=200\ $GeV 
are reported.
The cross section is generally consistent with
next-to-leading order perturbative QCD calculations.
The analyzing power is small at $x_F$ below about 0.3, and becomes 
positive and large at higher $x_F$, similar to the trend in data at 
$\sqrt{s}\le 20$ GeV.
The analyzing power is in qualitative agreement with perturbative QCD 
model expectations.
This is the first significant spin result seen for particles produced
with $p_T > 1\ $GeV/c at a polarized proton collider.
\end{abstract}

\maketitle
An early qualitative expectation from perturbative Quantum 
Chromodynamics (pQCD) was that the chiral properties of the theory 
would make the analyzing power for inclusive particle production be 
very small \cite{kane}.  
The analyzing power ($A_N$) is the azimuthal asymmetry in particle 
yields from a transversely polarized beam incident on an unpolarized 
target.
Earlier experiments studied polarized proton collisions ($p_\uparrow + p$) 
at center-of-mass energies $\sqrt{s}\le 20$ GeV and measured $A_N$ for 
pion production at moderate transverse momentum 
($0.5\le p_T\le 2.0\ $GeV/c).
Contrary to the naive expectation, $A_N$ was found to be $20-40\%$ 
for pions produced at large values of Feynman-$x$ ($x_F=2\, p_L/\sqrt{s}$, 
where $p_L$ is the longitudinal momentum of the pion) 
\cite{lowenergy,mediumenergy,E704,largeenergy}.
Similarly, elastic proton \cite{elastic} and recent semi-inclusive 
deep-inelastic lepton scattering experiments \cite{hermes,SMCconference} 
have reported transverse single-spin asymmetries which  
differ significantly from zero.  
These results have sparked substantial theoretical activity to understand 
transverse spin effects within the framework of pQCD \cite{review}.

Perturbative QCD calculations of pion production involve the convolution 
of parton distribution and fragmentation functions with a hard partonic 
interaction.
The reliability of calculations in the pQCD framework is expected to 
increase with $p_T$.
In this framework, forward $\pi$ production in $p+p$ collisions is 
dominated by scattering of a valence quark in one proton from a soft
gluon in the other. 
At large pseudorapidities ($\eta$) and $\sqrt{s}\le 20\ $ GeV, there may 
be significant contributions to particle production from soft hadronic 
processes collectively known as beam fragmentation.  
At a collider, $\sqrt{s}$ is significantly larger, leading to the
expectation that the origin of forward pions will shift towards 
collisions of the partonic constituents of the proton, consistent 
with the PYTHIA Monte Carlo generator \cite{PYTHIA}.
Measurements of the cross section for forward pion production
are important to establish that pQCD is a suitable framework for
treating polarization observables in these kinematics.

Different mechanisms have been identified in the pQCD framework by which 
one might expect transverse spin 
effects 
\cite{collins,sivers,efremov,anselminocollins,anselminosivers,qiusterman,koike}, 
all of which may contribute to some degree.  
With only data at $\sqrt{s}\le 20\ $GeV for comparison, these models are 
not well constrained.  
Despite this, the models have been extrapolated by an order of 
magnitude in $\sqrt{s}$ and approximately a factor of 2 in $p_T$, and all
predict that sizable transverse spin effects will persist at 
$\sqrt{s}=200\ $GeV.  
This Letter addresses the question if $A_N$ is sizable at 
$\sqrt{s}=200\ $GeV, as predicted by these models.
We present measurements of the cross section and $A_N$ for the 
production of forward $\pi^0$ mesons having $p_T > 1\ $GeV/c from 
$p_\uparrow +p$ collisions at $\sqrt{s}=200\ $GeV.

Data were collected by the STAR experiment (Solenoid Tracker at RHIC)
at the Brookhaven National Laboratory Relativistic Heavy Ion Collider 
(RHIC) in January 2002.
RHIC is the first polarized proton collider.
Polarization is produced by optical pumping
of an atomic-beam source \cite{anatoli} and is partially
preserved through an accelerator complex to reach RHIC \cite{bunce}.  
In RHIC, a pair of helical dipole magnets in each ring serves as the 
first use of full ``Siberian snakes'' \cite{siberian} in a 
high-energy accelerator to preserve polarization
during beam acceleration.  
The stable spin axis of the RHIC rings is vertical.  
Beam bunches crossed
the STAR interaction region (IR) every 213\ nsec.  
The polarization direction alternated between up and down for successive
bunches of one beam and after every two bunches of the other beam.  
Data were sorted according to the spin direction of the beam 
corresponding to positive $x_F$ pion production.
Summing all bunches in the other beam resulted in 
negligible remnant polarization.  
Typical luminosities were 
$10^{30}\ $cm$^{-2}$sec$^{-1}$,
and the integrated luminosity was $150\ $nb$^{-1}$ for these 
data.

The average beam polarization for each fill, $P_{beam}$, was 
determined using a Coulomb-Nuclear Interference (CNI) polarimeter 
located in RHIC \cite{kurita,spinka}.  
At $24.3\ $GeV, the RHIC injection energy, the analyzing power of the 
CNI reaction is $A^{CNI}_N=0.0133\pm 0.0041$ \cite{osamu,cni}, and 
can be used to deduce the absolute polarization of the proton beam.  
However, at $100\ $GeV, the beam energy used for RHIC collisions, 
$A^{CNI}_N$ has not yet been measured.  
The CNI asymmetries measured at injection and collision energies 
were nearly equal for many fills.  
Since the beam acceleration process is unlikely to increase 
$P_{beam}$, this suggests that $A^{CNI}_N$ at $100\ $GeV is no smaller
than at $24.3\ $GeV.
For the present analysis, we assume there is no change in $A^{CNI}_N$ 
between these two energies, giving an average value of 
$\langle P_{beam}\rangle =0.16$.

A prototype forward $\pi^0$ detector (PFPD) was installed at STAR
$750\ $cm from the IR to identify $\pi^0$ mesons.
At this time, STAR does not have the capability to identify large
rapidity charged pions.
The PFPD consisted of a Pb-scintillator sampling 
calorimeter \cite{pEEMCnim}, placed with its edge 
$\approx30\ $cm left of the oncoming polarized proton beam 
(beam-left).  
The PFPD was 21 radiation lengths deep and subdivided into 
$4\times 3$ towers.  
To measure the transverse profiles of photon showers, the PFPD 
had a shower-maximum detector (SMD) approximately 5 radiation lengths deep,
comprising two orthogonal layers of $100\times 60$ 
scintillator strips spaced at $0.37\ $cm.
To address systematic errors associated with measuring 
left-right asymmetries with a single arm detector, an array 
of Pb-glass detectors with no SMD was placed to the right of the 
oncoming polarized proton beam (beam-right). 
Similar arrays were placed above and below the vertically polarized
beam, where no asymmetries are expected.

The luminosity was measured at STAR using beam-beam counters (BBC)
\cite{joanna} composed of segmented scintillator annuli mounted around 
the beam at
longitudinal positions $z=\pm370\ $cm, spanning $3.3 < |\eta| <
5.0$.  Proton collision events were identified by requiring the
coincidence of at least one BBC segment fore and aft of the
IR.  Absolute luminosity was determined by measuring the
transverse size of the colliding beams and the number of protons
colliding at STAR.  
The cross section measured for the BBC coincidence condition is 
$26.1\pm 0.2(\mathrm{stat.})\pm 1.8(\mathrm{syst.})\ $mb
\cite{starpaper}, consistent with simulation \cite{PYTHIA,GEANT}.
The BBC observes $87\pm 8\%$ of the inelastic, non-singly diffractive 
cross section.

All forward calorimeters were read out when the energy deposited in 
any one calorimeter was greater than that from an electron of 
$\sim 15\ $GeV.
The BBC coincidence requirement was imposed to 
select $p+p$ collisions.

The asymmetry measured at beam-left is
\begin{equation}
   P_{beam} A_N = \frac{N_+ - R N_-}{N_+ + R N_-}.
\end{equation}
The number of $\pi^0$ mesons detected when the beam spin vector was
oriented up (down) is $N_{+(-)}$.  The spin-dependent relative
luminosity $(R={\mathcal L_+}/{\mathcal L_-}\approx 1.15)$ was measured 
with the BBC.
Background contributions to $R$ were reduced by increasing the 
coincidence requirements to at least two BBC segments on each side of 
STAR.  
The systematic errors on $R$, primarily coming from the change in 
$R$ when the background is corrected, are of the order of $10^{-3}$ 
\cite{joanna} and are a factor of 10 to 20 smaller than $P_{beam} A_N$ 
measured with the PFPD.

\begin{figure}
  \includegraphics[width=7.2cm,clip]{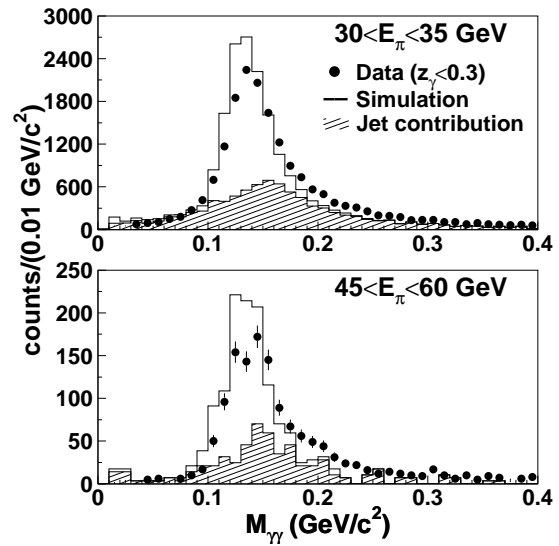}
  \caption{\label{fig:mass} Uncorrected spectra of the diphoton
  invariant mass in two energy bins.  
  The points are data with statistical errors.  
  The open histograms are reconstructed simulation events, 
  normalized to equal area.  
  The hatched histograms are used to correct
  the cross section.  }
\end{figure}

Neutral pions are reconstructed utilizing the formula
$M_{\gamma\gamma}=E_\pi \sqrt{1-z^2_\gamma}\, \sin
         (\phi_{\gamma\gamma}/2) \approx E_{tot}
         \sqrt{1-z^2_\gamma}\,d_{\gamma\gamma}/2 z_{vtx}$
using events with at least two clusters in the SMD.
The energy of the leading $\pi^0$, $E_\pi$, is taken to be the total
energy deposited in all of the towers, $E_{tot}$.  The opening angle 
between the two photons,
$\phi_{\gamma\gamma}$, is determined by $z_{vtx}$, the
distance between the collision vertex and the PFPD, and the
separation of the photons at the detector,
$d_{\gamma\gamma}$.  Both $d_{\gamma\gamma}$ and 
$z_{\gamma}=|E_{\gamma 1}-E_{\gamma 2}|/(E_{\gamma 1}+E_{\gamma 2})$ 
are measured by an analysis of the energy deposited in the strips of 
the SMD.  
The value of $d_{\gamma\gamma}$ is determined from the fitted centroids 
of the peaks, while $z_\gamma$ is derived from the ratio of the 
fitted areas under the peaks.  
A fiducial volume is defined by requiring the SMD peaks to be more than
12 strips from the detector edge.
Figure~\ref{fig:mass} shows the $M_{\gamma\gamma}$ spectra 
for two energy bins.
The mass resolution is $20\ $MeV/c$^2$ (RMS) for $15<E_{tot}<80$ GeV, 
limited by the measurement of $\phi_{\gamma\gamma}$.  
The centroid of the $\pi^0$ peak is used to determine the calibration
for each tower for each fill to an accuracy of the order of $1\%$.  
The calibration is found to have negligible dependence on energy or 
spin-state.

The $\pi^0$ detection efficiency is determined in a matrix of $E_\pi$
and $\eta$ from a Monte Carlo (MC) simulation of $p+p$ collisions 
\cite{PYTHIA} and the detector response \cite{GEANT}.  
The open histograms in Fig.~\ref{fig:mass} are MC
events which undergo the same reconstruction and selection as the data.  
The MC matches the data well for several variables,
including $p_T$, $E_{tot}$, and $\eta$.  
The $\pi^0$ detection efficiency is dominated by the geometrical 
acceptance of the calorimeter.

The $\pi^0$ sample is distorted by coincident particles from
the jets containing them.
The PFPD is about one hadronic interaction length deep.  
When two photons from $\pi^0$ decay overlap with 
other particles, the PFPD response to the other particles tends to 
increase $E_{tot}$ relative to $E_\pi$ and broaden the 
$\phi_{\gamma\gamma}$ resolution.
This results in a broad $M_{\gamma\gamma}$ distribution peaked at a 
value larger than $M_\pi$.  
The average value of $E_{tot}$ is about 3\ GeV larger than 
$E_\pi$, independent of $E_\pi$.  
MC events with $|E_{tot}-E_\pi| > 2\ $GeV are shown as the 
hatched histograms in Fig.~\ref{fig:mass}.  
Events with only one photon from $\pi^0$ decay plus other particles 
exist predominantly at small $M_{\gamma\gamma}$, and are suppressed by 
requiring $z_\gamma < 0.3$.
The $E_\pi$-dependent systematic error in the cross section is 
about $20\%$, dominated by the jet correction.
The MC simulation includes $\pi^0$ events from forward jets.
The uncertainty includes the difference when these effects are 
explicitly corrected in both the data and the simulation, and in 
neither.

Non-collision background is suppressed to the level of $1\%$ by
requiring the coincidence from the BBC in the offline analysis.
Following our simulations, the cross section is corrected by $10\%$ to
account for the bias introduced by the BBC coincidence condition.  
Hadronic background comprising events with no leading
$\pi^0$ in the acceptance of the calorimeter is predominantly at small
$M_{\gamma\gamma}$, and is reduced by constraining $z_\gamma$.  
The hadronic background amounts to about $2\%$ of the yield
underneath the $\pi^0$ peak at $0.09<M_{\gamma\gamma}<0.22\ $GeV/c$^2$.  

\begin{figure}
  \includegraphics[width=7.2cm,clip]{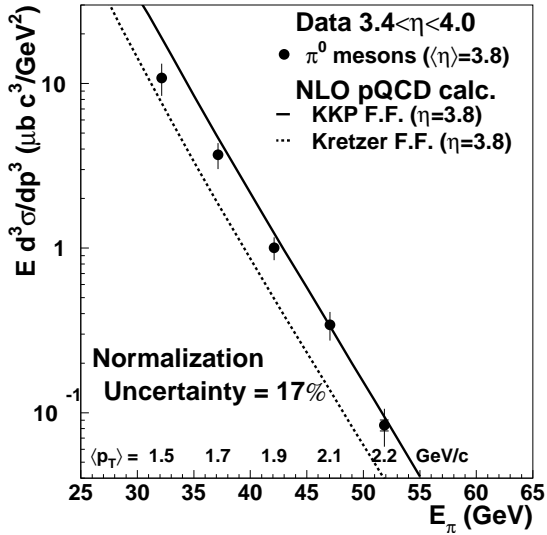}
  \caption{\label{fig:crosssec} Inclusive $\pi^0$ production cross 
    section versus leading $\pi^0$ energy ($E_\pi$).  
    The average transverse momentum ($\langle p_T\rangle$) is correlated 
    with $E_\pi$, as the PFPD was at a fixed pseudorapidity ($\eta$).
    The inner error bars are statistical, and are smaller 
    than the symbols for most points.  The outer error bars combine
    these with the $E_\pi$-dependent systematic errors. 
    The curves are NLO pQCD calculations evaluated at 
    $\eta=3.8$ \cite{vogelsang,kniehl,kretzer}.  }
\end{figure}

The inclusive $\pi^0$ production cross section for 
$30 < E_\pi < 55\ $GeV in 5\ GeV bins is presented in
Fig.~\ref{fig:crosssec}.
Data with $3.4 < \eta < 4.0$ were selected, giving 
$\langle\eta\rangle =3.8$ independent of $E_{\pi}$; in this range the 
detector efficiency is well understood.  
The dominant contributions to the normalization error come from 
knowledge of the absolute transverse position of the detector (10\%), 
the absolute luminosity determination (8\%), and the model 
dependence of the BBC efficiency (8\%).  
The data are plotted at the average $E_{\pi}$ of the bin.

The curves on the plot are next-to-leading order (NLO) pQCD
calculations \cite{vogelsang} evaluated at $\eta=3.8$, using the
CTEQ6M \cite{cteq} parton distribution functions and equal 
renormalization and factorization scales of $p_T$.  
The NLO pQCD calculations are in general consistent with the data, in 
contrast to midrapidity $\pi^0$ data at lower $\sqrt{s}$ \cite{aurenche}.
The solid line uses the ``Kniehl-Kramer-P\"{o}tter'' (KKP) set of
fragmentation functions (F.\,F.) \cite{kniehl}, while the dashed line uses the
``Kretzer'' set \cite{kretzer}.  The difference between the two
reflects uncertainties in the F.\,F.\ at these
kinematics.  
At the chosen scale, KKP tends to agree with 
the data better than Kretzer, consistent with midrapidity $\pi^0$ data 
at $\sqrt{s}=200\ $GeV \cite{phenix}.

\begin{figure}
  \includegraphics[width=7.2cm,clip]{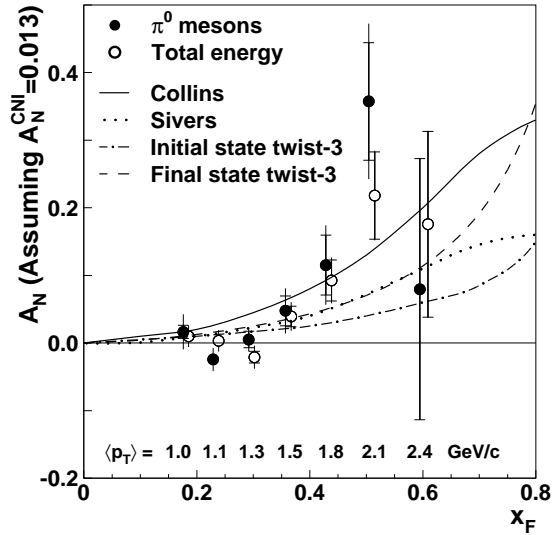}
  \caption{\label{fig:anvsxf} Analyzing powers versus Feynman-$x$ 
    ($x_F$).
    The average transverse momentum ($\langle p_T\rangle$) is correlated 
    with $x_F$.
    The solid points are for identified $\pi^0$ mesons.  
    The open points are for the total energy ($E_{tot}$), shifted by 
    $x_F+0.01$. 
    The inner error bars are statistical, and the outer combine these 
    with the point-to-point systematic errors.  
    The curves are from pQCD models evaluated at $p_T=1.5\ $GeV/c 
    \cite{anselminocollins,anselminosivers,qiusterman,koike}.  
    The $A_N$ values are proportional to $A_N^{CNI}$, assumed to 
    be 0.013 at 100 GeV.}
\end{figure}

The analyzing power is presented in
Fig.~\ref{fig:anvsxf}, plotted versus 
$2\, \langle E_{tot}\rangle /\sqrt{s}\approx x_F$.
The solid points are for $\pi^0$ mesons from 
$3.3 < \eta < 4.1$ and $0.07 < M_{\gamma\gamma} < 0.3\ $GeV/c$^2$, 
with $x_F$-dependent constraints on $z_\gamma$ to minimize
background.  
The open points are based solely on $E_{tot}$ in 
the PFPD without SMD analysis:  neither fiducial volume constraints nor 
$\pi^0$ identification.
The agreement between the solid and open points indicates $A_N$ is not 
sensitive to the analysis used to identify $\pi^0$ mesons.  
This is consistent with simulations showing that $95\%$ of events 
with at least 25\ GeV deposited in the PFPD come from photons, $95\%$
of which are daughters from $\pi^0$ decay.
The $A_N$ seen at beam-right with the Pb-glass array is similar to
that seen at beam-left with the PFPD, while $A_N$ for the Pb-glass 
above and below the beam is consistent with zero, as expected.  
The largest $x_F$-dependent systematic error arises from 
comparison of the beam-left and beam-right data.  
The average $A_N(x_F)$ is computed using both,
and an uncertainty is assigned to bring
$A_N(x_F)$ (shown in Fig.~\ref{fig:anvsxf}) within 1 standard deviation 
of the average.

The curves on the plot are predictions from the pQCD models, fitted 
to data at $\sqrt{s}=20\ $GeV, extrapolated to $\sqrt{s}=200\ $GeV
and evaluated at $p_T=1.5\
$GeV/c \cite{anselminocollins,anselminosivers,qiusterman,koike}.  One
model attributes single-spin effects to the convolution of the
transversity distribution function with a spin-dependent Collins
fragmentation function \cite{anselminocollins}.
The Sivers model adds explicit spin-dependent $k_T$ dependence to the 
parton distribution functions \cite{anselminosivers}.
Other models ascribe the effects to twist-3 parton correlations in the
initial or final state \cite{qiusterman,koike}.  
The data are qualitatively consistent with all of these predictions.

The trend of $A_N$ at lower $\sqrt{s}$ is to increase
from zero beginning at a value of $x_F$ which depends on
$\sqrt{s}$ \cite{largeenergy}.  
The significance of the increase for these data is $4.7\,\sigma$ 
(including statistical and point-to-point systematic errors)
from a linear fit to the open circles in Fig.~\ref{fig:anvsxf} for 
$x_F > 0.27$, with $\chi^2=0.9$ for 3 degrees of freedom.  
This is the first significant spin result seen for particles
with $p_T > 1\ $GeV/c at a polarized proton collider.

In summary, high energy $\pi^0$ mesons have been observed 
at forward angles from
$p_\uparrow +p$ collisions at $\sqrt{s}=200\ $GeV.
The differential cross section is, in general, consistent with 
NLO pQCD calculations.
The analyzing power is small at $x_F$ below about 0.3, and 
becomes positive and large at higher $x_F$, similar to the trend 
observed in fixed-target data at $\sqrt{s}\le 20$ GeV.
The analyzing power at $\sqrt{s}=200\ $GeV is in qualitative agreement 
with pQCD model predictions.  
Higher precision measurements of $A_N$ as a function of both $x_F$ and 
$p_T$ may help to differentiate among the models.  
Future measurements may attempt to determine the Collins fragmentation 
function in $p_\uparrow +p$ collisions, as well as to look at jet 
production and Drell-Yan scattering to isolate potential contributions 
to transverse spin effects.

We thank the RHIC Operations Group and RCF at BNL, and the NERSC
Center at LBNL for their support. This work was supported in part by
the HENP Divisions of the Office of Science of the U.S.  DOE; the
U.S. NSF; the BMBF of Germany; IN2P3, RA, RPL, and EMN of France;
EPSRC of the United Kingdom; FAPESP of Brazil; the Russian Ministry of
Science and Technology; the Ministry of Education and the NNSFC of
China; Grant Agency of the Czech Republic, DAE, DST, and CSIR of the
Government of India; the Swiss NSF.


\begin{thebibliography}{10}
  \bibitem{kane} G.\,L.~Kane, J.~Pumplin, and W.~Repko, Phys.\ Rev.\
    Lett.\ {\bf 41}, 1689 (1978).

  \bibitem{lowenergy} R.\,D.~Klem {\it et al.}, Phys.\ Rev.\ Lett.\
    {\bf 36}, 929 (1976); W.\,H.~Dragoset {\it et al.}, Phys.\ Rev.\
    D\ {\bf 18}, 3939 (1978).

  \bibitem{mediumenergy} S.~Saroff {\it et al.}, Phys.\ Rev.\ Lett.\
    {\bf 64}, 995 (1990); B.\,E.~Bonner {\it et al.}, Phys.\ Rev.\ D\
    {\bf 41}, 13 (1990).

  \bibitem{E704} B.\,E.~Bonner {\it et al.}, Phys.\ Rev.\ Lett.\ {\bf
    61}, 1918 (1988); A.~Bravar {\it et al.}, {\it ibid.}\ {\bf 77},
    2626 (1996); D.\,L.~Adams {\it et al.}, Phys.\ Lett.\ B {\bf 261},
    201 (1991); {\bf 264}, 462 (1991); Z.\ Phys.\ C {\bf 56}, 181 
    (1992).

  \bibitem{largeenergy} K.~Krueger {\it et al.}, Phys.\ Lett.\ B\ {\bf
    459}, 412 (1999); C.\,E.~Allgower {\it et al.}, Phys.\ Rev.\ D\
    {\bf 65}, 092008 (2002).

  \bibitem{elastic} P.\,R.~Cameron {\it et al.}, Phys.\ Rev.\ D\
    {\bf 32}, 3070 (1985);
    D.\,G.~Crabb {\it et al.}, Phys.\ Rev.\ Lett.\ {\bf 65}, 3241 
    (1990).
  
  \bibitem{hermes} A.\,Airapetian {\it et al.}, Phys.\ Rev.\ Lett.\
    {\bf 84}, 4047 (2000); Phys.\ Lett.\ B {\bf 535}, 85 (2002); {\bf
    562}, 182 (2003).

  \bibitem{SMCconference} A.~Bravar {\it et al.}, Nucl. Phys.
    Proc. Suppl. {\bf 79}, 520 (1999).

  \bibitem{review} For a review, see V.~Barone, A.~Drago,
    and P.\,G.~Ratcliffe, Phys.\ Rep.\ {\bf 359}, 1 (2002).

  \bibitem{PYTHIA} T.~Sj\"{o}strand, Comp. Phys. Commun. {\bf 82}, 74
    (1994).

  \bibitem{collins} J.~Collins, Nucl.\ Phys.\ {\bf B396}, 161 (1993).

  \bibitem{sivers} D.~Sivers, Phys.\ Rev.\ D\ {\bf 41}, 83
      (1990); {\bf 43} 261 (1991).

  \bibitem{efremov} A.~Efremov and O.~Teryaev, Phys.\ Lett.\
      {\bf 150B}, 383 (1985).

  \bibitem{anselminocollins} M.~Anselmino, M.~Boglione, and F.~Murgia, 
    Phys.\ Rev.\ D\
    {\bf 60}, 054027 (1999); M.~Boglione and E.~Leader, Phys.\ Rev.\
    D\ {\bf 61}, 114001 (2000).

  \bibitem{anselminosivers} M.~Anselmino, M.~Boglione, and
    F.~Murgia, Phys.\ Lett.\ B\ {\bf 362}, 164 (1995); M.~Anselmino
    and F.~Murgia, {\it ibid.}\ {\bf 442}, 470 (1998); U. D'Alesio 
    and F. Murgia, AIP Conf.\ Proc.\ {\bf 675} 469 (2003).

  \bibitem{qiusterman} J.~Qiu and G.~Sterman, Phys.\ Rev.\ D\ {\bf
    59}, 014004 (1998).

  \bibitem{koike} Y.~Koike, AIP Conf.\ Proc.\ {\bf 675}, 449 (2003).

  \bibitem{anatoli} A.~Zelenski {\it et al.}, in 
    {\it Proceedings of the Particle Accelerator Conference}, 
    (IEEE, New York, 1999) p.\ 106.
                  
  \bibitem{bunce} G.~Bunce {\it et al.}, Ann.\ Rev.\ Nucl.\ Part.\ Sci.\ 
               {\bf 50}, 525 (2000); 
     H.~Huang {\it et al.}, Phys.\ Rev.\ Lett.\ {\bf 73}, 2982 (1994); 
     M.~Bai {\it et al.}, {\it ibid.}\ {\bf 80}, 4673 (1998).

  \bibitem{siberian} 
    Ya.\,S.~Derbenev {\it et al.}, Part.\ Acc.\ {\bf 8}, 115 (1978).

  \bibitem{kurita} I.\,G.~Alekseev {\it et al.}, AIP Conf.\ Proc.\
     {\bf 675}, 812 (2003).

  \bibitem{spinka} H.~Spinka, AIP Conf.\ Proc.\ {\bf 675}, 807 (2003).

  \bibitem{osamu} O.~Jinnouchi {\it et al.}, AIP Conf.\ Proc.\ 
     {\bf 675}, 817 (2003).

  \bibitem{cni} J.~Tojo {\it et al.}, Phys.\ Rev.\ Lett.\ {\bf 89},
    052302 (2002).

  \bibitem{pEEMCnim} C.~Allgower {\it et al.}, Nucl.\ Instr.\ 
    Meth.\ {\bf A499}, 740 (2003).

  \bibitem{joanna} J.~Kiryluk, AIP Conf.\ Proc.\ {\bf 675}, 424 
     (2003).

  \bibitem{starpaper} J.~Adams {\it et al.}, Phys.\ Rev.\ Lett.\ 
    {\bf 91}, 172302 (2003);
    A.~Drees and Z.~Xu, in {\it Proceedings of the Particle 
    Accelerator Conference} (IEEE, Chicago, IL, 2001) p.\ 3120.

  \bibitem{GEANT} GEANT 3.21, CERN program library.

  \bibitem{vogelsang} F.~Aversa {\it et al.}, Nucl.\ Phys.\ 
    {\bf B327}, 105 (1989);
    B.~Jager {\it et al.}, Phys.\ Rev.\ D\ {\bf 67}, 054005 (2003); 
    D.~de~Florian, {\it ibid.}\ {\bf 67}, 054004 (2003).

  \bibitem{kniehl} B.\,A.~Kniehl {\it et al.}, Nucl.\ Phys.\ {\bf B597},
    337 (2001).

  \bibitem{kretzer} S.~Kretzer, Phys.\ Rev.\ D\ {\bf 62}, 054001 (2000).
 
  \bibitem{cteq} J.~Pumplin {\it et al.}, J.\ High Energy Phys.\ 
    {\bf 07} (2002) 012.

  \bibitem{aurenche} P.~Aurenche {\it et al.}, Eur.\ Phys.\ J.\ C\
    {\bf 13}, 347 (2000).

  \bibitem{phenix} S.\,S.~Adler {\it et al.}, 
       Phys.\ Rev.\ Lett.\ {\bf 91}, 241803 (2003).

\end{thebibliography}
\end{document}